\documentclass[useAMS,usenatbib,onecolumn]{mnras}

\usepackage{dcolumn}
\usepackage{graphicx}
\usepackage{url}
\usepackage{color}
\usepackage{mathtools}
\usepackage{threeparttable}

\usepackage{longtable}

\usepackage{booktabs}

\graphicspath{ {eps/} }

\newcommand{\cm}{cm$^{-1}$}

\newcommand{\ai}{\textit{ab initio}}
\newcommand{\Ai}{\textit{Ab initio}}

\newcommand{\Duo}{{\sc Duo}}

\title[ExoMol XXIII: Line lists for PO and PS]{ExoMol molecular line lists  - XXIII:  spectra of PO and PS}

\date{\today}
\author[Prajapat  et al.]{Laxmi Prajapat$^{1}$, Pawel Jagoda$^{1}$, Lorenzo Lodi$^{1}$,  Maire N. Gorman$^{1,2}$,
\newauthor Sergei N. Yurchenko$^{1}$ and Jonathan Tennyson$^{1}$\thanks{Email: j.tennyson@ucl.ac.uk}  \\
$^{1}$ Department of Physics and Astronomy, University College London, London WC1E 6BT,
UK\\
$^{2}$ Department of Physics, Aberystwyth University, Penglais, Aberystwyth, Ceredigion, UK, SY23 3BZ}
\date{Accepted XXXX. Received XXXX; in original form XXXX}
\pubyear{2017}

\pagerange{\pageref{firstpage}--\pageref{lastpage}}
\begin{document}
\maketitle
\begin{abstract}

Comprehensive
line lists for phosphorus monoxide ($^{31}$P$^{16}$O) and phosphorus monosulphide ($^{31}$P$^{32}$S) in
their $X$~$^2\Pi$ electronic ground state are presented.
The line lists are based on new
\ai\ potential energy (PEC), spin-orbit (SOC) and dipole moment
(DMC) curves computed using the MRCI$+$Q$-$r method with aug-cc-pwCV5Z and
aug-cc-pV5Z basis sets. The nuclear motion equations (i.e. the rovibronic Schr\"odinger equations for each molecule) are solved using the program \Duo.
The PECs and SOCs are refined in least-squares fits
to available experimental data.  Partition functions, $Q(T)$, are computed
up to $T=$ 5000~K, the range of validity of the line lists.
These line lists are the most comprehensive available for either molecule.
The characteristically sharp peak of the $Q$-branches from the spin-orbit split
components give
useful diagnostics for both PO and PS in spectra at infrared wavelengths.
These line lists should
prove useful for analysing observations and setting up models of environments
such as brown dwarfs, low-mass
stars,  O-rich circumstellar regions and potentially  for
exoplanetary retrievals.
Since PS is yet to be detected in space, the role of the two lowest excited
electronic states ($a$~$^4\Pi$ and $B$~$^2\Pi$) are also considered.
An approximate line list for the PS $X$ -- $B$ electronic transition, which predicts
a number of sharp vibrational bands in the near ultraviolet, is also presented.
The line lists are available from the CDS \url{http://cdsarc.u-strasbg.fr} and ExoMol
 \url{www.exomol.com} databases
\end{abstract}


\label{firstpage}

\section{Introduction}

Several phosphorus-containing molecules have been discovered around evolved stars including PN, HCP, CP and PO \citep{08MiHaTe.PN, 07TeWoZi.PO, 13DeKaPa.PO} and PH$_{3}$ has been detected in the circumstellar envelope of IRC$+$10216 \citep{14AgCeDe.PH3}. Other phosphorous-containing species including PS are yet to be detected \citep{13DeKaPa.PO}. Models suggest that a variety of phosphorous-bearing species become important in the atmospheres of low-mass stars, brown dwarfs and giant exoplanets at elevated temperatures \citep{06ViLoFe.PH3}. PN has been observed in the 3 mm region towards the low-mass star forming region of L1157 \cite{11YaTaSa.PN, 12YaTaWa.PN}.

Additionally, being a primal biogenic element found in all living systems,
phosphorus is hence essential to life on Earth \citep{97MaHeOr}.
Phosphorus is present in nucleic acids, several proteins, and is a
fundamental component of the adenosine triphosphate (ATP) molecule,
which is accountable for energy transfer in cells. P-containing molecules
are thought to provide important biomarkers in the early Earth \citep{13LiSuCh}
and such  molecules could play a similar role
in exoplanets.
In addition, the atmospheres of the recently characterised hot rocky planets, or lava planets,
are likely to contain a whole range of unusual small molecules \citep{jt693}.
In this work, performed as part of the ExoMol project \citep{jt528}, we concentrate
on providing comprehensive line lists for
two open shell diatomic species: PO and PS.
Oxygen and sulphur belong to the same group in the periodic table and, as a result, PO and PS
have similar electronic structures (e.g.both have a $X$~$^2\Pi$ ground electronic state) and
their spectra show many analogies.

After a number of failed attempts \citep{87MaFeBe.PO,01MaChaxx.PO,01DiPaEl.PO},
phosphorus monoxide, PO, has been detected in a number of locations in space.
The original detection, by \citet{07TeWoZi.PO}, was in the oxygen-rich, red Supergiant Star VY Canis Majoris and used
microwave emissions near 240 and 284 GHz (7.2 and 8.5 cm$^{-1}$). Subsequently PO has been observed
in the wind of the oxygen-rich AGB star IK Tauri \citep{13DeKaPa.PO}, and in star-forming
regions \citep{16RiFoBe.PO,16LeVaVi.PO}. In a number of these locations,
PO appears to occur with similar abundance
to the closed shell molecule PN for which an ExoMol line list has already been constructed \citep{jt590}.

As of yet there are no observations of  phosphorus monosulphide, PS, in space \citep{13DeKaPa.PO}. A systematic  attempt at
its astronomical detection was performed  by  \citet{88OhYaSa.PS}  using the 45m telescope of the Nobeyama Radio Observatory (NRO) and
six distinct objects.
Local Thermodynamic Equilibrium (LTE) calculations by \citet{73Tsuji}
indicate that PS should be the major P-bearing molecule in oxygen-rich circumstellar envelopes for
temperatures below 2000 K.

Available line lists for both PO and PS appear to be extremely limited.
Long-wavelength transition frequencies are available for both species in the JPL database \citep{jpl},
but the transition intensities are all based on assumed or estimated values for the permanent dipole moments.
A more up-to-date long-wavelength line list for PO is given by the Cologne
Database for Molecular Spectroscopy (CDMS) \citep{CDMS}.

There have been several studies which aimed to obtain line
frequencies and spectroscopic constants of PO from experimental and theoretical
analyses of its spectrum
\citep{03MePaMa.PO,03MoOuDe.PO,12SuWaSh.PO,13LiShSu.PO,99SpHaxx.PO,78GhVexxa.PO,02BaBoDe.PO,83BuKaHi.PO,95Qian.PO,99deBrou.PO,81RaReRa.PO,88KaYaSa.PO}.
 A review
of the experimental work on PO prior to 1999 is given by
\citet{99deBrou.PO}, while \citet{13LiShSu.PO} provides a
more recent summary of \ai\ studies.
In addition to work on the spectrum of PO in its $X$~$^{2}\Pi$ ground electronic state, there have also been extensive experimental
studies of its excited electronic states. Early work on observed transitions
is summarised by  \citet{79HeHuxx.book}; Huber and Herzberg list in their compilation for PO
13 electronic states up to about 56~000~\cm\ (six $^{2}\Sigma^+$ states, one $^{2}\Sigma^-$ state, three
$^{2}\Pi$ states, two $^{2}\Delta$ state and one $^{4}\Sigma^-$ state).
Additional excited electronic states, including more quartet and sextet states, have been considered theoretically
\citep{88KaYaSa.PO,99SpHaxx.PO,99deBrou.PO,00deBrxx.PO,03MoOuDe.PO,03MePaMa.PO,12SuWaSh.PO,13LiShSu.PO}.

PS was observed in the laboratory for the first time by
\citet{55DrMixx.PS} who detected two band systems corresponding to the
$C$~$^{2}\Sigma\rightarrow$ $X$~$^{2}\Pi$ and $B$~$^{2}\Pi\rightarrow$ $X$~$^{2}\Pi$
electronic transitions, with wavelength ranges 2700-3100~\AA\
and 4200-6000~\AA\ respectively.
Since then a limited set of experiments on PS have followed
\citep{69NaSuxx.PS,71NaBaxx.PS,78JePaxx.PS,79BaDiNa.PS,87LiBaWr.PS,88KaHiOh.PS,88OhYaSa.PS,93GuWaGu.PS} with the most recent
being the study of a submillimeter-wave rotational spectrum
by \citet{99KlKlWi.PS}. Several of these studies are considered further below.
There have also been various theoretical studies conducted on the
ground and electronic states of PS
\citep{87BrGrxx.PS,88KaBrGr.PS,92KaGrxx.PS,98MoOuDe.PS,02Kalche.PS,12YaFrHo.PS}, including an MRCI study of the lowest 16 molecular terms by  \citet{12YaFrHo.PS}.

The aim of this work is to produce molecular line lists for $^{31}$P$^{16}$O and $^{31}$P$^{32}$S applicable for a large range of temperatures.

\section{Method}

In the theoretical approach adopted by our group \citep{jt475} the
computation of a line list for the molecule of interest
requires constructing potential energy
curves (PECs), dipole moment curves (DMCs), spin-orbit couplings
(SOCs) and, if necessary, other couplings such
as angular momentum, spin-spin  and spin-rotation \citep{jt632}.
These are then used to solve the relative nuclear-motion Schr\"{o}dinger equation,
thus producing frequencies and intensities for the transitions of
interest.

\subsection{\textit{Ab initio} electronic structure data}

\subsubsection{PO}

\begin{figure}
\center
\includegraphics[width=0.45\textwidth]{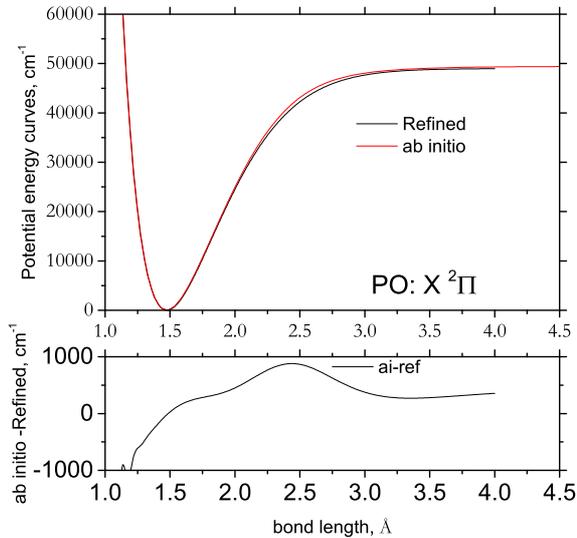}
\caption{Comparison of \ai\ (MRCI$+$Q$-$r/aug-cc-pwCV5Z) and refined PECs of PO. }
\label{f:PO_ai}
\end{figure}

While there have been several \ai\
studies of PO's many electronic states which yielded total energies
and spectroscopic constants,  none of them supply the data required on
PECs, SOCs and DMCs for the construction of a line list. \Ai\ curves were therefore computed
 using \textsc{molpro}
\citep{MOLPRO}. The chosen technique was MRCI$+$Q$-$r (internally contracted
multi-reference configuration interaction with renormalized Davidson
correction; the `relaxed reference' energy was used)
with the aug-cc-wCV5Z basis set. The calculation also included a relativistic
correction curve computed as the expectation value of the
mass-velocity plus one-electron Darwin operator (MVD1). The
\ai\ PEC was computed up to a nuclear seperation of ~4.5 \AA\ and can be seen in Fig. \ref{f:PO_ai}.

The \ai\ SOC was obtained using the same
level of theory.  Fig.~\ref{f:PO_SOC}
shows the \ai\ SO curve, as well as
the refined curve (see below). The equilibrium \ai\ SO value (106.4~\cm) is in
reasonable agreement with the empirical SO constant ($A_{\rm SO}/2 =
112.1$~\cm) from \citet{83BuKaHi.PO} and with the \ai\ value 112.6~\cm\ of
\citet{13LiShSu.PO}. Our refined equilibrium SO value is 112.1~\cm.

\begin{figure}
\includegraphics[width=0.65\textwidth]{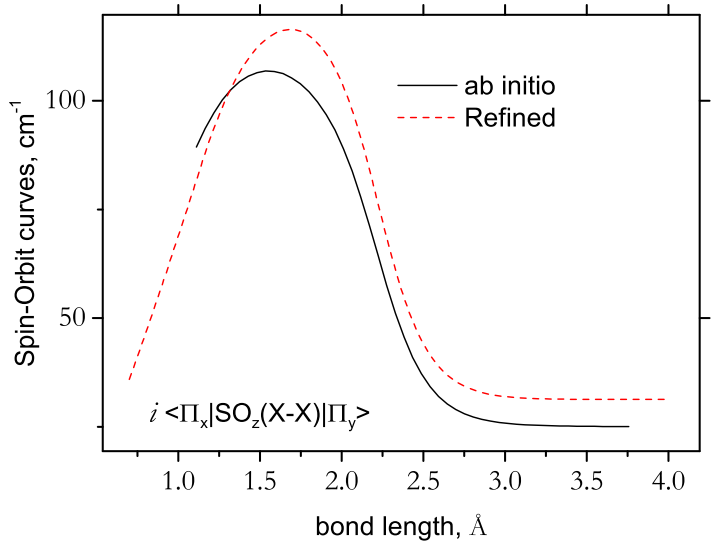}
\caption{Comparison of PO spin-orbit curves. Solid line: \ai\  curve; red dashed line: empirical curve produced using the morphing approach. }
\label{f:PO_SOC}
\end{figure}

Our value of the equilibrium bond length $r_{\rm e}=1.482$~\AA\ obtained from the PEC of the $X$~$^{2}\Pi$ state is in
reasonable agreement with the most recent experimentally determined value
$r_{\rm e}=1.475637355(10)$~\AA\ \citep{02BaBoDe.PO}. The dissociation
value $D_{\rm e}=48997.9$ cm$^{-1}$ also compares well with the
value 48980~\cm\ (6.073~eV) estimated by \citet{81RaReRa.PO}.
This \ai\ PEC thus provides a suitable starting point for empirical refinement of the $X$~$^{2}\Pi$ state.

At the start of this work there was no DMC for the $X$~$^{2}\Pi$ state
of PO available in the literature with instead the majority of \ai\
studies focusing on the myriad of low-lying PECs of PO.
\citet{03MoOuDe.PO} computed a value for the dipole at equilibrium and
recently \citet{16AnDeBo.PO} computed curves as part of their study on
the formation of PO by radiative association.

The experimental data is limited to the $\mu_0$
value by \citet{81RaReRa.PO}.  Therefore, it was decided
to use the highest level of \ai\ theory in this work to produce a suitable DMC
for the line list calculation. Similar to the final \ai\ PEC, the \ai\ DMC was
calculated using the MRCI$+$Q$-$r/aug-cc-pwCV5Z; it was calculated as the
derivative of the MRCI$+$Q$-$r energy with respect to an external electric field along
the internuclear axis for vanishing field strength \citep{jt475}.

\begin{figure}
\center
\includegraphics[width=0.45\textwidth]{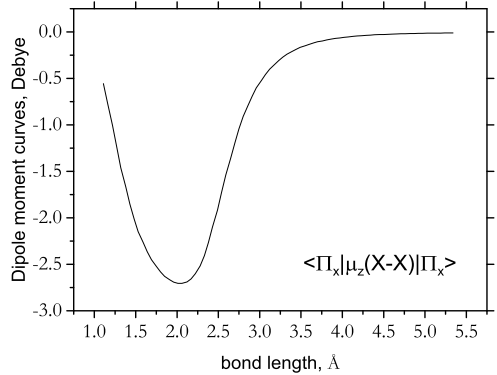}
\caption{\Ai\ dipole moment curve (DMC) for the ground state of PO.}
\label{f:PO_DMC}
\end{figure}

The  dipole moment generated for the $X$~$^{2}\Pi$ state of PO is shown in Fig. \ref{f:PO_DMC}. Its value at the equilibrium
bond length is  1.998 D, which is in reasonable agreement with
1.88$\pm0.07$~D \citep{88KaYaSa.PO} (also adopted by CDMS \citep{CDMS}) and therefore provides an adequate choice in the final calculation of the line list.
These values are much larger than the value of 1.0 D assumed in the JPL line list \citep{jpl}.

\subsubsection{PS}

\begin{figure}
\centering
\includegraphics[width=12cm]{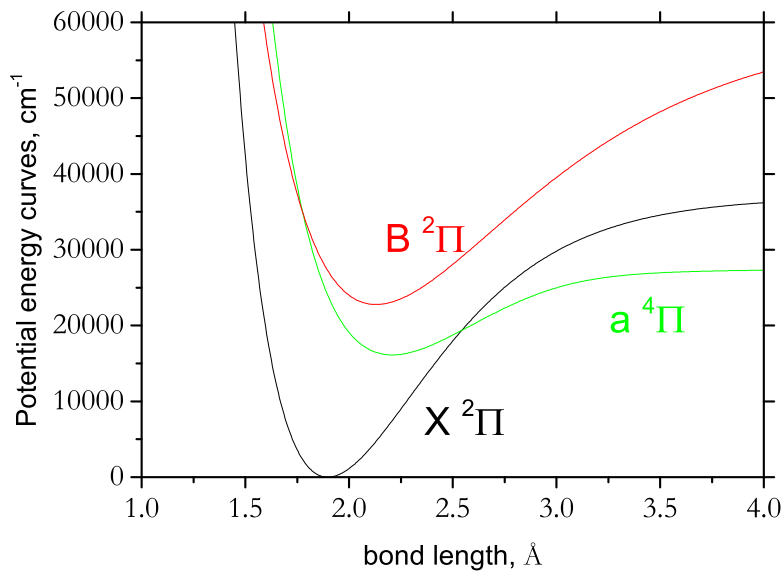}
\caption{Refined PECs for the $X$~$^{2}\Pi$, $B$~$^{2}\Pi$ electronic states and the \ai\ PEC for the $a$~$^{4}\Pi$ electronic state of PS.}
\label{f:PS_PEC:X-A-a}
\end{figure}

\textit{Ab initio} curves for the ground  ($X$~$^{2}\Pi$),
and three excited  ($a$~$^{4}\Pi$, $B$~$^{2}\Pi$ and
$^{4}\Sigma^{-}$) electronic states of PS were generated using \textsc{molpro} and the MRCI$+$Q$-$r method with the aug-cc-pwCV5Z basis sets used for
the ground state and aug-cc-pV5Z basis sets used for the excited states.
The \ai\ PEC of the
$a$~$^4\Pi$ state and the refined PECs for the $X$~$^2\Pi$ and $B$~$^2\Pi$ states considered in this
paper are shown in Fig.~\ref{f:PS_PEC:X-A-a}.
The dissociation limit obtained by our calculation for the $X$~$^{2}\Pi$ electronic state appears to be close to the
estimated $D_{\rm e}$ = 36600 \cm\ (438$\pm$10 kJ mol$^{-1}$) of
\citet{73DrMySz.PS}, although we did not perform calculation at large enough
bond lengths to quote an accurate value. Our equilibrium dipole moment of the $X$ state is 0.523~D,
which can be compared to the complete basis set (CBS) extrapolation value by \citet{13MuWoxx.PS} of 0.565~D.

\begin{figure}
\centering
\includegraphics[width=12cm]{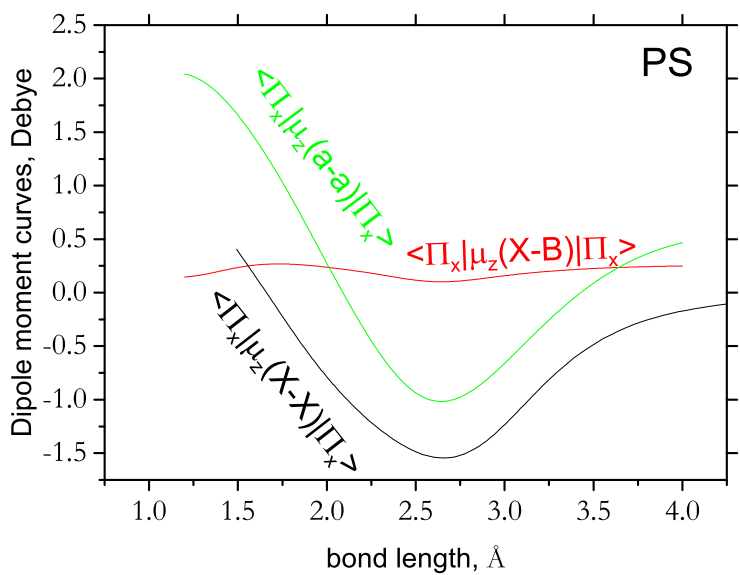}
\caption{\Ai\ dipole moment curves of PS which were represented analytically.}
\label{f:PS_DMC}
\end{figure}

The \ai\ DMCs for the ground $X$~$^{2}\Pi$ electronic state, excited $a$~$^{4}\Pi$ electronic state and for the $X$~$^{2}\Pi$--$B$~$^{2}\Pi$ electronic transition considered in this work for PS are shown in Fig.~\ref{f:PS_DMC}. In order to reduce the numerical noise when computing the line-strengths using the \Duo\ program, we followed the recommendation of \citet{16MeMeSt} and represented analytically the \ai\ DMCs (denoted by  $\mu(r)$). The following expansion with a damped-coordinate was employed:
\begin{equation}\label{e:dipole}
  \mu(r) = (1-\xi) \sum_{n \ge 0} d_n z^n   + d_{\infty} \, \xi ,
\end{equation}
where $\xi$ is the \v{S}urkus variable \citep{84SuRaBo.method} 
\begin{equation}
\label{e:surkus}
\xi=\frac{r^{p}-r^{p}_{\mathrm{ref}}}{r^{p}+r^{p}_{\rm ref }}  
\end{equation}
and $z$ is given by
$$
z = (r-r_{\rm ref})\, e^{-\beta_2 (r-r_{\rm ref})^2-\beta_4 (r - r_{\rm ref})^4}.
$$
Here $p$ is an empirical parameter, $r_{\rm ref}$ is a reference
position equal to $r_{\rm e}$ by default, $d_n$ are the expansion
parameters, $d_{\infty}$ is the value of the dipole at $r\to \infty$
and $\beta_2$ and $\beta_4$ are damping factors.  These parameters
defining the dipole moment expansion for three \ai\ DMCs considered in
this work for PS are given in supplementary material as a \Duo\ input
file, while the functional form is now a part of the \Duo\ program.

\begin{figure}
\centering
\includegraphics[width=12cm]{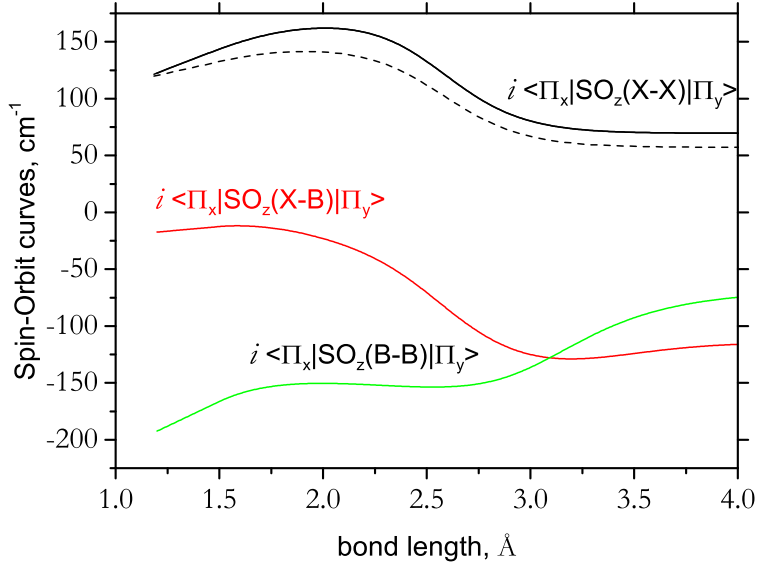}
\caption{Spin-orbit curves of PS, \ai\ (solid) and refined (dashed).} 
\label{f:PS_SOCs}
\end{figure}

The SOCs were computed using the aug-cc-pVDZ basis set (valence only calculations) to speed up their calculation (SOC are not expected to be very
sensitive to the level of theory used, see \citet{jt589}) and are shown in Fig.~\ref{f:PS_SOCs}.
Our equilibrium SO \ai\ value  is 141.4~\cm\ and after refinement is 160.9~\cm.
The analogous experimental effective SO constant $A_{SO}/2$ was determined to be 161.0~\cm\ by \citet{78JePaxx.PS}.

\subsection{Nuclear motion calculations}

The nuclear motion calculations were performed using the code \Duo\
\citep{jt609}, which provides a variational solution to the nuclear
motion problem and can account for virtually any coupling between the
PECs of the molecule under study.  A review of the theory on which
\Duo\ is based can be found in \citet{jt632}.  The necessary curves
can be computed using \ai\ electronic methods or by fitting to
experimental data. Our general strategy \citep{jt511,jt693}, which is
followed here, is to use spectroscopically determined PECs and
couplings, since these cannot be computed with sufficient accuracy by
\ai\ methods \citep{jt623}. Conversely, experience has shown that \ai\
DMCs give results which are very reliable \citep{jt573} and can give
intensities which are competitive in accuracy with the most precise
laboratory measurements \citep{jt613}. In general, the nuclear
motion problem can be solved with sufficient accuracy that the
quality of the underlying \ai\ curves is the main source of error in the
calculations.

\subsection{Refinement using experimental data}

\subsubsection{PO}

Table \ref{t:PO_expt} lists the sources used in refining the PO $X$~$^{2}\Pi$ PEC. Most of the 241 lines used correspond to pure rotational transitions observed
in microwave and far-infrared studies, including hot bands $\Delta v = 0$ and $v'\le 7$. The only source of IR data is the laser spectroscopy study by \citet{95Qian.PO} where 50 fundamental transitions ($J\le 21.5$) were reported. \citet{83BuKaHi.PO} and \citet{02BaBoDe.PO}  resolved the
hyperfine structure within each observed rotational transition. As
hyperfine splitting is beyond the scope of this work,
frequencies resulting from such transitions
were averaged over for each rotational transition.

We also included lower accuracy experimentally-derived vibrational energies by \citet{75VeSixx.PO} with $T_{v}$ with $v\leq 11$. The availability of these vibrational terms values provided an important constraint for extrapolation of higher vibrational levels than $v=1$.

It is easier to start the \Duo\ refinement by fitting to energies, not transition frequencies.
To this end, an `experimentally derived' set of energy levels of PO was produced using the program PGOPHER \citep{PGOPHER}, using the experimentally derived spectroscopic constants for the ground ($v=0$) and first excited ($v=1$) vibrational states, as determined in \citet{95Qian.PO}. These energies (up to $J = 26.5$) were combined with the vibrational energies by \citet{75VeSixx.PO}.

\begin{table}
\caption{Experimental sources used in the empirical refinement of the PO
 PEC.}
\begin{tabular}{lllrcc}
\hline
Source & Method & Transitions & lines & Range / cm$^{-1}$ & Uncertainty
cm$^{-1}$\tabularnewline
\hline
\citet{02BaBoDe.PO} & Millimeter wave &$\Delta v=0$, $v=0:7, J\leq10.5$
&167  & 4.98 - 15.39 & 0.01\tabularnewline
\citet{95Qian.PO} & Microwave, IR  & $v=1\leftarrow0$, $J\leq21.5$,
&46 & 1188.12 - 1245.12 & 0.0005\tabularnewline
\citet{83BuKaHi.PO} & mid-IR  & $\Delta v=1$, $v=0-1, J\leq25.5$,
&28  & 2.1226 - 1254 & 0.0005 \tabularnewline
\citet{75VeSixx.PO} & UV & $T_{v}$ for $v=0:11$ &&$E_{v}=0-12700.05$ & 0.02\tabularnewline
\hline
\end{tabular}\label{t:PO_expt}
\end{table}

\subsubsection{PS}

There is little laboratory data available on the $X$~$^{2}\Pi$ state of PS.
\citet{55DrMixx.PS}, \citet{69NaSuxx.PS}, \citet{71NaBaxx.PS}, and \citet{79BaDiNa.PS} reported
vibronic heads only and did not provide any information that can be used to refine this state.
\citet{88KaHiOh.PS}, \citet{88OhYaSa.PS} and \citet{99KlKlWi.PS} give millimeter wave spectra
which provide information on the rotational levels and spin-orbit splitting between the states.
\citet{88OhYaSa.PS} and \citet{99KlKlWi.PS} provide hyperfine-resolved transition frequencies which
were unresolved
by averaging the frequencies of matching transitions with the same
e/f, ${J}'$, ${J}''$ and $\Omega$ values where $\Omega$ is the projection of the total angular momentum.
The e/f parity was not provided
for $\Omega=\frac{3}{2}$ in \citet{88OhYaSa.PS} and \citet{88KaHiOh.PS}; it was
assumed to be unresolved and were duplicated for later use.
\citet{88KaHiOh.PS} provides data on the vibrational fundamental.
In addition, we used the program \textsc{pgopher}
to derive a set of PS energy levels from the spectroscopic constants reported in the
experimental paper by \citet{99KlKlWi.PS}. This gave a total of 316 energies for
 $J$ ranging from 0.5 to 39.5 split between the $v=0$ and $v=1$ state. For the vibrationally excited states we added the vibrational energies reconstructed from the vibrational $\Delta v= 1$ separations
 reported by \citet{78JePaxx.PS} from analysis of the $B$~$^{2}\Pi$--$X$~$^{2}\Pi$ system. The energies coincided with consecutive vibrational transitions from ${v=0-9}$.

To make all data internally consistent, the experimental e/f parities for the
lower state were converted into $+/-$ parity \citep{75BrHoHu.parity} using the standard relations:
\begin{equation}
\mathrm{e:\:}(-1)^{\mathrm{J-\frac{1}{2}}}
\end{equation}
and
\begin{equation}
\mathrm{f:\:}(-1)^{\mathrm{J+\frac{1}{2}}}.
\end{equation}
The selection
rule $+\leftrightarrow-$ was used to determine the parity
for the upper state. The frequency and quantum numbers were repeated
for both $+$ and $-$ whenever the parities were unavailable
in the experimental data (all in the case of the $^{2}\Pi_{\frac{3}{2}}$
sub-state).  $\Sigma'$ and $\Sigma''$ values (projections of the electronic spin on the molecular axis) were derived from $|\Omega|$ usually provided
in experimental literature as $|\Omega|=|\Lambda \pm \Sigma'|$ and by matching the corresponding parity. Here $\Lambda=\pm 1$ is the projection of the electronic angular momentum and $\Omega$ is the projection of the total angular momentum on the molecular axis.

\subsection{Fitting with \Duo}

\Duo\ offers a range of analytical functions
for modelling PECs. Owing to its previous success in producing accurate
PECs, it was decided to use the Extended Morse Oscillator (EMO) function
\citep{EMO} to obtain the final $X$~$^{2}\Pi$ PECs for PO and PS denoted by $V(r)$. The function is written:
\begin{equation}
V(r)=D_{\rm e}\left[1-\exp\left(-{\sum_{k=0}^{N}}\beta_{k}\xi^{k}(r-r_{\rm e})\right)\right]^{2},
\label{e:EMO}
\end{equation}
where $\xi$ is the \v{S}urkus variable, see Eq.~(\ref{e:surkus}). Here $\beta_{k}$ is an empirical parameter whose value (along with the parameter $p$) can be
derived through refinement to experimental data. It is important to note that $V(+\infty)=D_{\rm e}$, as long as both $p$ and $\underset{k}{\sum}\beta_{k}$
are set greater than zero.

\Duo\ also allows the SOC to be refined simultaneously with the PEC.
An  \ai\ curve $F_{\rm ai}(r)$ can be scaled using the morphing approach
so that the empirical curve, $F(r)$, is given by
\begin{equation}
F(r)=H(r)F_{\rm ai}(r)
\end{equation}
with
\begin{equation}
H(r)=\left[(\mathrm{1}-\xi)\, {\sum_{k=0}^{N}}\,\beta_{k}\, \xi^{k}+\xi \, t_{\infty}\right]
\label{e:H(R)}
\end{equation}
where $H(r)$ is the morphing function in terms of the \v{S}urkus variable ($\xi$), $t_{\infty}$
is the value of the morphing function as $r\rightarrow\infty$, and
$\beta_{k}$ is the morphing expansion coefficient \citep{jt609}. The morphing approach
was used to refine the \ai\ SOCs.  As
there were no available \ai\ curves
for spin-rotation (SR) and $\Lambda$-doubling effects, the functional form  $H(r)$ was applied to both couplings directly.

The final \Duo\ input files for both PO and PS are given as part of the supplementary
material. These files contain the various curves as well as the parameters used to
run \Duo.

\subsubsection{PO}

Experimental values of PO for dissociation $D_{\rm e}$ \citep{81RaReRa.PO} and equilibrium bond length
$r_{\rm e}$ \citep{02BaBoDe.PO} were
held fixed until the final stages of the refinement process. The first four $\beta_k$ expansion
coefficients from Eq.~(\ref{e:EMO}) were varied by fitting to experimentally derived energies of PO obtained using \textsc{pgopher} until a satisfactory fit was achieved. At this point the empirical energies were
replaced by the actual, measured frequencies for comparison and further refinement. As the reference frequencies only include pure rotational
and fundamental absorption lines, to increase constraints on the higher
vibrational transitions, the difference between the vibrational energies
\citep{75VeSixx.PO} relative to the ground level ($v=0$) were retained.

At the final stage of refinement, two additional terms were introduced
to account for spin-rotation and any additional $\Lambda$-doubling
effects to further minimise Obs.$-$Calc. The resulting empirical PEC
and SOC are shown in Fig. \ref{f:PO_ai} and Fig. \ref{f:PO_SOC}, respectively: these are plotted with the
respective \ai\ curves for comparison. The final parameters as well as the corresponding curves are given as part of the \Duo\ input files in supplementary data.


The accuracy of
the fit ranges from 0.001 cm$^{-1}$ for purely
rotational transitions up to 0.05 cm$^{-1}$ for
vibrational transitions, producing a root mean square (RMS) error of 0.014 cm$^{-1}$. The residuals are illustrated in Fig.~\ref{f:PS_Obs-Calc}, where the vibronic bands are indicated.
Table~\ref{t:vib} illustrates the Obs.-Calc. residuals for the
vibrational excitation's of PO and PS. In both cases the empirical
term values are of limited accuracy \citep{75VeSixx.PO,78JePaxx.PS},
especially for the PS values from \citet{78JePaxx.PS}.
While the target accuracy of the PEC was achieved for low vibrational
levels, having access to more experimental data for
higher levels would help to verify the accuracy of extrapolated energy
levels in the line list.

\begin{table}
\caption{Obs.-Calc. residuals for PO and PS vibrational term values in \cm. The experimentally derived
energies of PO and PS are from \protect\citet{75VeSixx.PO} and \protect\citet{78JePaxx.PS}, respectively.}
\label{t:vib}
\begin{tabular}{rrrrcrrr}
\hline\hline
& \multicolumn{3}{c}{PO} & \multicolumn{3}{c}{PS} \\
\hline
$v$  &  Obs.       &  Calc.   &  Obs.-Calc &   & Obs.       &  Calc.   &  Obs.-Calc \\
\hline
  1  &  1220.161  &  1220.216  &   -0.055 &   &   1461.7 &   1461.4 &      0.3 \\
  2  &   2427.31  &  2427.291  &    0.019 &   &   2183.4 &   2183.2 &      0.2 \\
  3  &   3621.29  &  3621.242  &    0.04 &   &   2899.0 &   2899.0 &      0.0 \\
  4  &   4802.12  &  4802.078  &    0.04 &   &   3608.9 &   3608.8 &      0.1 \\
  5  &   5969.84  &  5969.799  &    0.04 &   &   4312.3 &   4312.6 &     -0.3 \\
  6  &   7124.41  &  7124.398  &    0.01 &   &   5010.2 &   5010.2 &      0.0 \\
  7  &   8265.85  &  8265.865  &   -0.01 &   &   5701.8 &   5701.7 &      0.1 \\
  8  &   9394.14  &  9394.184  &   -0.04 &   &          &          &          \\
  9  &   10509.30  &  10509.33  &   -0.03 &   &          &          &          \\
 10  &   11611.30  &  11611.29  &    0.01 &   &          &          &          \\
 11  &  12700.05  &  12700.03  &    0.02 &   &          &          &          \\
\hline
\end{tabular}
\end{table}

\subsubsection{PS}

Again the PEC was fitted using an EMO.
Values of $r_{\rm e}$ and
${D_{\rm e}}$ were kept fixed to their spectroscopic
values, $r_{\rm e}$ = 1.89775~\AA\
and $D_{\rm e}= 37~004$ \cm.
The \ai\ PEC and SOC curves for the ground state of
PS were used as the starting point in the fits.

The experimental set for the fit comprised of the ${v}=0-9$  energies
from \citet{78JePaxx.PS}, and the reconstructed energies from
\textsc{pgopher} for ${J}$ up to 39.5. Although
the experimental data was limited, \Duo\ is able to extrapolate
energies to higher values for the $v$ and $J$ quantum numbers.
The empirical PEC is compared with the \ai\ one in
Fig. \ref{f:PO_ai}.

Morphing was used
to refine the SOC.
The experimental value for $r_{\rm e}$ was  used as the reference
expansion point. ${t_{\infty}}$ represents the asymptote
for the morphing function (${r\rightarrow\infty}$), and it
equals unity in this case \citep{jt589}. Two morphing expansion parameters,
${A_{0}}$ and ${A}_{1}$, were included in the fit
because the Obs.-Calc was reduced even further, however no significant changes
were made when fitting to the expansion parameters for ${n}>1$.

\begin{figure}
\centering
\includegraphics[width=11cm]{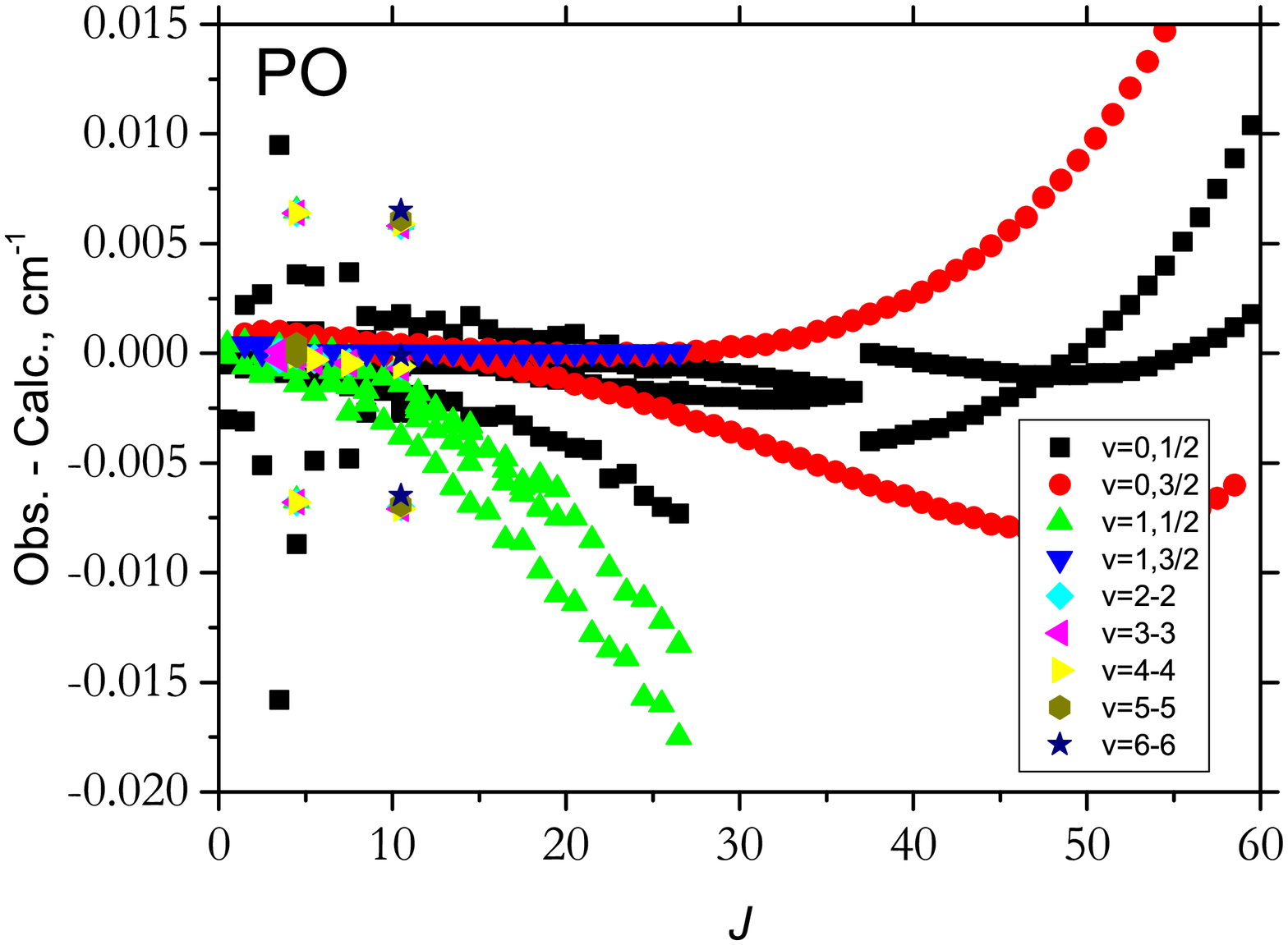}
\includegraphics[width=11cm]{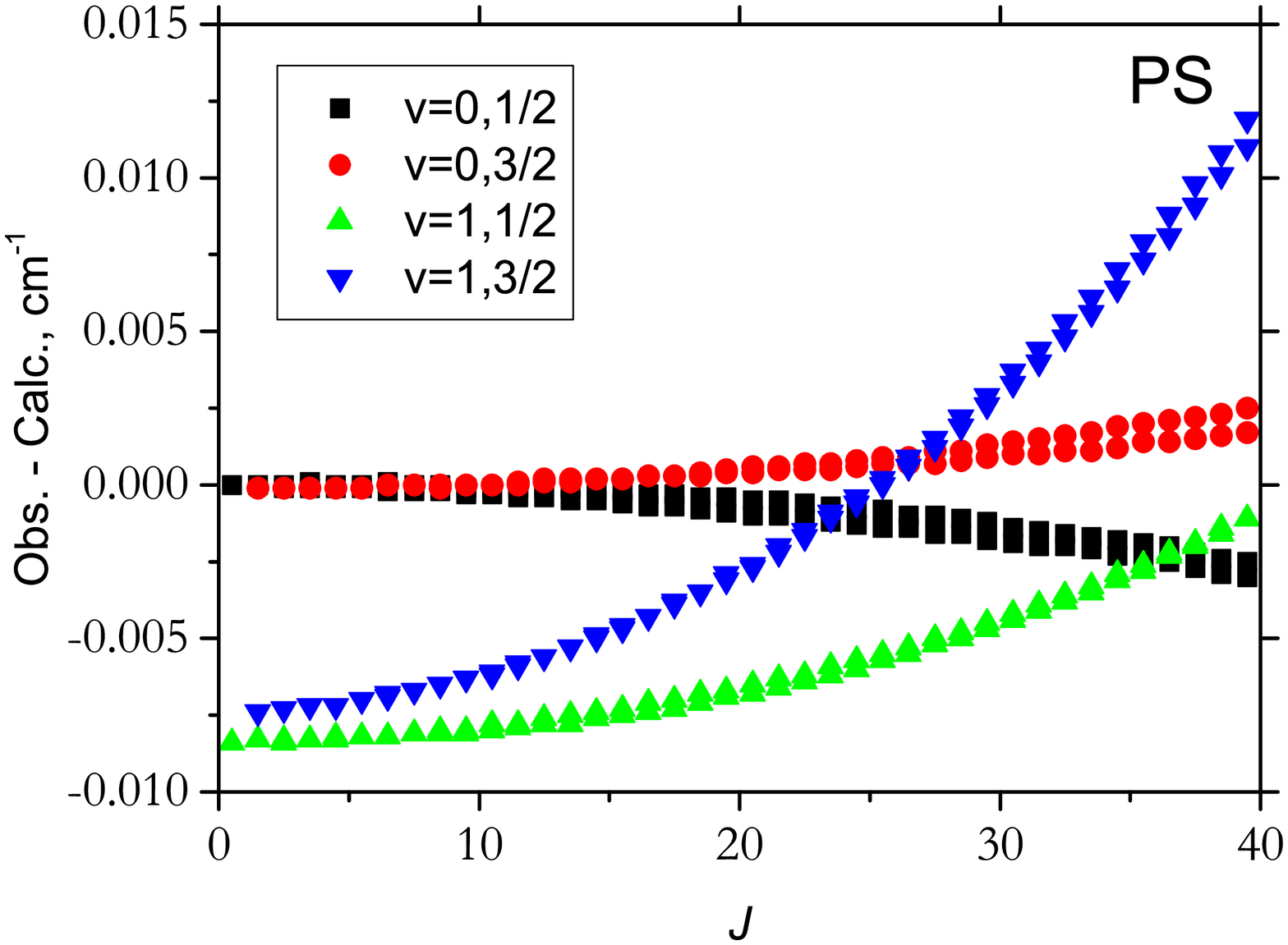}
\caption{Observed minus Calculated (\cm) PO and PS energy term levels and line positions after refinement. The discontinuity in the PS comparison is probably because the CDMS data switches from three to two hyperfine components  at $J = 36.5$, which we average to one hyperfine-free value.  }
\label{f:PS_Obs-Calc}
\end{figure}

Residues of the fit (Obs.-Calc.) for both molecules
are plotted in Fig~\ref{f:PS_Obs-Calc} for all the data used in the fitting except
for the less accurate vibrational energies of \citet{75VeSixx.PO} and
\citet{78JePaxx.PS}, which are collected in Table~\ref{t:vib}.
The residuals build distinct vibronic $v,\Omega$  patterns and diverge somewhat at higher $J$
indicating a deficiency in our model. One of the possible sources of error is the $\Lambda$-factor for $\Pi$ states,
which, when properly modelled, should originate from the
electronic angular momentum coupling to $\Sigma$ states \citep{79BrMexx.methods}.
The majority of the points are situated near the zero line.

The RMS error of the fit was 0.026~\cm\ for all 316 experimental energies and 0.006~\cm\ excluding the lower accuracy vibrational data of  \citet{78JePaxx.PS}.

For PS we have decided to include the two lowest excited electronic states  $a$~$^{4}\Pi$ and $B$~$^2\Pi$ by using the corresponding \ai\ MRCI$+$Q$-$r/aug-cc-pV5Z curves. In order not to destroy the accuracy of the refined model of the ground electronic state of PS, we omitted the couplings between these two states with the $X$ state. The $B$~$^{2}\Pi$ state PEC was refined by fitting to $J=0.5$ and $J=1.5$ rovibronic energies of this state, which we derived using parameters ($T_v$, $r_{\rm e}$  and $B_{v}$) of \citet{78JePaxx.PS} (RMS error is 0.41~\cm).




\section{Line List calculations}

\subsection{PO ($X$) and PS ($X$, $B$, $a$) line lists }

Line lists generated using \Duo\  are comprised of
a states file and a transitions file, with extensions \texttt{.states}
and \texttt{.trans}, respectively \citep{jt631}. The \texttt{.states} file includes
the running number $n$, energy term values (\cm), total
statistical weight, lifetime \citep{jt624}, $g$-Land\'{e} factors \citep{jt656} and corresponding quantum numbers. The \texttt{.trans}
file contains running numbers for the upper and lower levels, as well
as the Einstein-A coefficients \citep{jt609}.

The maximum vibrational and rotational quantum numbers, ${v_{max}}$ and ${J_{max}}$ are identified using the dissociation energy, ${D_{0}}$. These numbers are given for both
PO and PS in Table~\ref{t:stats}.
The \Duo\ integration range was chosen as $r=[0.7,4.0]$ \AA\ for PO and  $r=[1.2,4.0]$ \AA\ for PS, and the respective grids comprised 501 points in conjunction with the DVR sinc method. The PO line list includes the electronic state $X^2\Pi$, while the PS line list consists of transitions between the lowest three electronic states of $X^2\Pi$, $B^2\Pi$ and $a^4\Pi$. To reduce the size of the line list, for PS the lower state energy threshold is reduced to 25,000~\cm, which should cover all thermal populations much higher than 5000~K.

As an additional safeguard against enhanced intensities for high overtones resulting from numerical noise,
we follow the procedure used by \citet{jt686} and use a dipole moment threshold of
$10^{-8}$~D.

\begin{figure}
\includegraphics[width=0.7\textwidth]{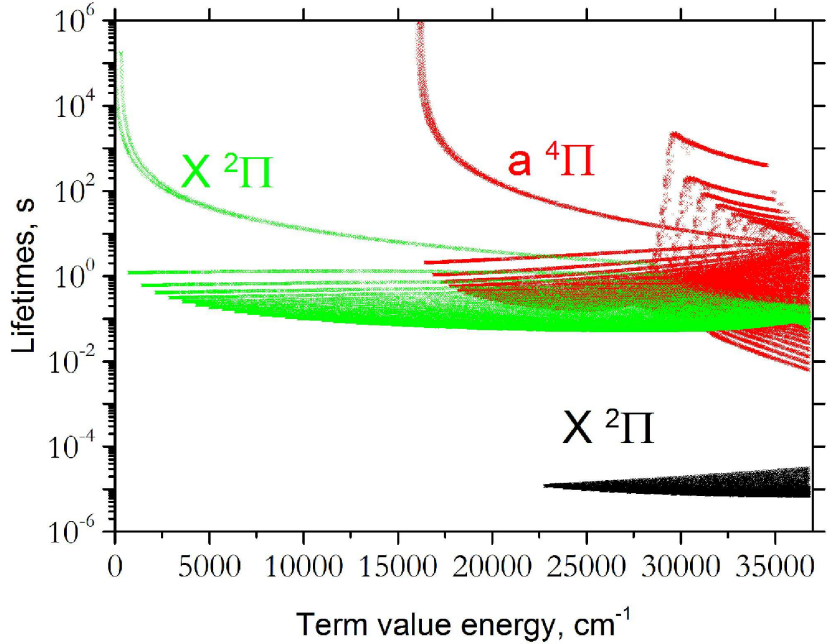}
\caption{Lifetimes of the three lower electronic states of PS computed using the PS line list.}
\label{f:lifetime}
\end{figure}

Lifetimes for PS molecule were computed using an extended line list covering all transitions with the lower/upper state energies below 37,000~\cm. They are shown in Fig.~\ref{f:lifetime}.

Table~\ref{t:stats} summarises the statistics for our PO and PS line lists.
Extracts from the \texttt{.states} and \texttt{.trans} files for PS
are given in Tables~\ref{t:PS_states} and \ref{t:PS_trans}, respectively.
Full tables for both PO and PS are available at \url{ftp://cdsarc.u-strasbg.fr/pub/cats/J/MNRAS/xxx/yy},
or \url{http://cdsarc.u-strasbg.fr/viz-bin/qcat?J/MNRAS//xxx/yy},
as well as the ExoMol website, \url{www.exomol.com}.
Note that the states files include lifetime for each individual state \citep{jt624}; these can be useful for a variety of issues including determining critical densities.

\begin{table}
\caption{Statistics for line lists for PO and PS}

\begin{tabular}{lrrrrr}
\hline\hline
& PO & PS \\
\hline
States ($v_{\rm max}$)& $X$(69)      & $X$(82),$B$(35),$a$(58)   \\
$J_{\rm max}$          & 234.5      & 320.5         \\
$\nu_{\rm max}$ (\cm)  & 12,000     & 37,000        \\
$E'_{\rm max}$ (\cm)   & 49,000     & 37,000        \\
$E''_{\rm max}$ (\cm)   & 37,000     & 25,000        \\
number of energies     & 43,148     & 225,997      \\
number of lines        & 2,096,289  & 30,394,544   \\
\hline
\end{tabular}
\label{t:stats}
\end{table}

\begin{table}
\caption{Extract from the \texttt{.states} file for $^{31}$P$^{32}$S.}
{\tt
\begin{tabular}{rrrrrrrrrrrrr}
\toprule
$n$ & $\tilde{E}$ & $g_{\rm tot}$ & $J$ & $\tau$ & $g$-Land\'{e} & $+/-$ &  e/f & State & $v$ & $\Lambda$ &$\Sigma$ & $\Omega$ \\
\hline
   1 &    0.000000    &   4    &  0.5  &  inf          &  -0.000767   &   +  &  e & X2Pi &   0 &   1 & -0.5 &  0.5 \\
   2 &  733.657367    &   4    &  0.5  &  1.2003E+00   &  -0.000767   &   +  &  e & X2Pi &   1 &   1 & -0.5 &  0.5 \\
   3 & 1461.408539    &   4    &  0.5  &  6.0697E-01   &  -0.000767   &   +  &  e & X2Pi &   2 &   1 & -0.5 &  0.5 \\
   4 & 2183.215742    &   4    &  0.5  &  4.0934E-01   &  -0.000767   &   +  &  e & X2Pi &   3 &   1 & -0.5 &  0.5 \\
   5 & 2899.040733    &   4    &  0.5  &  3.1065E-01   &  -0.000767   &   +  &  e & X2Pi &   4 &   1 & -0.5 &  0.5 \\
   6 & 3608.844801    &   4    &  0.5  &  2.5151E-01   &  -0.000767   &   +  &  e & X2Pi &   5 &   1 & -0.5 &  0.5 \\
   7 & 4312.588793    &   4    &  0.5  &  2.1216E-01   &  -0.000767   &   +  &  e & X2Pi &   6 &   1 & -0.5 &  0.5 \\
   8 & 5010.233137    &   4    &  0.5  &  1.8413E-01   &  -0.000767   &   +  &  e & X2Pi &   7 &   1 & -0.5 &  0.5 \\
   9 & 5701.737858    &   4    &  0.5  &  1.6315E-01   &  -0.000767   &   +  &  e & X2Pi &   8 &   1 & -0.5 &  0.5 \\
  10 & 6387.062584    &   4    &  0.5  &  1.4689E-01   &  -0.000767   &   +  &  e & X2Pi &   9 &   1 & -0.5 &  0.5 \\
  11 & 7066.166547    &   4    &  0.5  &  1.3393E-01   &  -0.000767   &   +  &  e & X2Pi &  10 &   1 & -0.5 &  0.5 \\
  12 & 7739.008581    &   4    &  0.5  &  1.2337E-01   &  -0.000767   &   +  &  e & X2Pi &  11 &   1 & -0.5 &  0.5 \\
  13 & 8405.547119    &   4    &  0.5  &  1.1461E-01   &  -0.000767   &   +  &  e & X2Pi &  12 &   1 & -0.5 &  0.5 \\
  14 & 9065.740185    &   4    &  0.5  &  1.0724E-01   &  -0.000767   &   +  &  e & X2Pi &  13 &   1 & -0.5 &  0.5 \\
  15 & 9719.545397    &   4    &  0.5  &  1.0095E-01   &  -0.000767   &   +  &  e & X2Pi &  14 &   1 & -0.5 &  0.5 \\
  16 &10366.919955    &   4    &  0.5  &  9.5535E-02   &  -0.000767   &   +  &  e & X2Pi &  15 &   1 & -0.5 &  0.5 \\
\bottomrule
\end{tabular}
}

\begin{tabular}{cll}
\\
             Column       &    Notation                 &      \\
\midrule
   1 &   $n$              &       Energy level reference number (row)    \\
   2 & $\tilde{E}$        &       Term value (in \cm) \\
   3 &  $g_{\rm tot}$     &       Total degeneracy = $g_{ns} J(J+1)$ with $g_{ns}=2$  \\
   4 &  $J$               &       Rotational quantum number    \\
   5 & $\tau$ & Radiative lifetime (s) \\
   6 & $g$ & Land\'{e} factors  \\
   7 & $+/-$ & Total parity  \\
   8 & $e/f$ & Rotationless parity \\
   9 & State & Electronic state \\
  10 & $v$ & State vibrational quantum number \\
  11 &  $\Lambda$ &   Projection of the electronic angular momentum \\
  12 & $\Sigma$ &  Projection of the electronic spin \\
  13 & $\Omega$ &   $\Omega=\Lambda+\Sigma$ (projection of the total angular momentum) \\
\bottomrule
\end{tabular}
\label{t:PS_states}

\end{table}

\begin{table}
\center
\tt
\caption{Extract of the first 15 lines from the PS \textsf{.trans} file. Identification numbers \textit{f} and \textit{i} for
upper (final) and lower (initial) levels, respectively, Einstein-A coefficients denoted by \textit{A} (s$^{-1}$)
and transition frequencies $\nu$ (cm$^{-1}$).}
\begin{tabular}{rrrr} \hline\hline
\multicolumn{1}{c}{\textit{i}} & \multicolumn{1}{c}{\textit{f}} & \multicolumn{1}{c}{\textit{A}} & \multicolumn{1}{c}{$\nu$} \\
\hline
       98398     &    98831  &  5.1886E-17&        0.012535  \\
       97958     &    97524  &  4.6732E-17&        0.012571  \\
      101886     &   102318  &  7.6872E-17&        0.012588  \\
      103188     &   102758  &  8.6033E-17&        0.012611  \\
      100140     &   100573  &  6.2520E-17&        0.012617  \\
       99701     &    99268  &  5.6651E-17&        0.012627  \\
      101445     &   101012  &  6.8893E-17&        0.012630  \\
       98396     &    98829  &  5.1177E-17&        0.012655  \\
      101884     &   102316  &  7.5904E-17&        0.012672  \\
      103186     &   102756  &  8.4859E-17&        0.012678  \\
       97956     &    97522  &  4.6028E-17&        0.012698  \\
      100138     &   100571  &  6.1750E-17&        0.012726  \\
      101443     &   101010  &  6.8055E-17&        0.012730  \\
\bottomrule
\end{tabular}
\label{t:PS_trans}
\end{table}

\subsection{Partition Functions}

Partition functions, $Q(T)$, for $^{31}$P$^{16}$O and $^{31}$P$^{32}$S
were computed using the program ExoCross \citep{jt698}, which was
also used to calculate absorption and emission cross-sections.  The
nuclear statistical weight of both species is ${g_{\rm ns}=2}$.  The
maximum temperature was set to 5000~K and partition functions were
determined in increments of 1~K.  Table \ref{t:pf} gives partition function values
at selected temperatures and a comparison with various studies.
The full tabulation is given in the supplementary material.

\begin{table}
\caption{Comparison of calculated partition function
$Q(T)$ with those of \protect\citet{84SaTaxx.partfunc},
\protect\citet{81Irwin.partfunc},
\protect\citet{16BaCoxx.partfunc}$^a$ and JPL \protect\citep{jpl}.}
\vspace*{-3mm}
\begin{threeparttable}
\begin{tabular}{lrrrrrr}
\toprule
&\multicolumn{1}{c}{$T$} & \multicolumn{1}{c}{This work} & \multicolumn{1}{c}{Irwin} & \multicolumn{1}{c}{Sauval \& Tatum} &\multicolumn{1}{c}{Barklem \&  Collet} &\multicolumn{1}{c}{JPL}  \\
\hline
PO\\
 &300    &   1544.45&            &  2094.29 &   2303.22  &1539.8440 \\
 &1000   &   7986.52& 7987.18    & 9182.62  &   9269.00  &          \\
 &2000   &  24522.37& 24517.88   & 26386.24 &   26532.2  &          \\
 &3000   &  50383.83& 50326.68   & 52435.51 &   53203.8  &          \\
 &4000   &  86046.26& 85818.70   & 88045.63 &   89775.2  &          \\
 &5000   & 132015.51& 131377.58  & 133966.53&   136893.4 &          \\
\hline
PS\\
&	300	&	3525.47	&	&	3381.35	&	3532.94	&	3522.0557\\
&	1000	&	23627.9	&	&	23710.1	&	23602.8	&	\\
&	2000	&	83950.66	&	&	82351.29	&	83383.6	&	\\
&	3000	&	183804.33	&	&	177908.26	&	181066.8	&	\\
&	4000	&	329835.22	&	&	313112.07	&	318202	&	\\
&	5000	&	538812.55	&	&	490668.67	&	497144	&	\\
\hline\hline
\end{tabular}
\begin{tablenotes}
\item[]
$^a$Partition function values from \citet{84SaTaxx.partfunc,81Irwin.partfunc,16BaCoxx.partfunc} are doubled to allow for nuclear spin
degeneracy and to bring them into line with the convention used here and by the
other cited sources.
\end{tablenotes}
\end{threeparttable}
\label{t:pf}
\end{table}

For PO our partition function is in good agreement with that of
\citet{81Irwin.partfunc} but agrees less well with \citet{84SaTaxx.partfunc} or the recent one of \citet{16BaCoxx.partfunc}, whose
values appear to be too high.
For PS, the agreement with \citet{16BaCoxx.partfunc} is altogether more satisfactory, although their values become a little too low
at the higher energies.

The partition functions from this study were represented by
the series expansion following the recommendation of \citet{jt263}
\begin{equation}
\log_{10}(Q)=\sum_{i=0}^{8} a_{i}(\log_{10}(T))^{i}.
\label{e:pffit}
\end{equation}
The nine expansion coefficients denoted by $a_{i}$ are collected as part of the supplementary material.
The fits are valid for temperatures up to 5000~K.

\section{Results}

\subsection{PO $X$~$^{2}\Pi$ state}

\begin{figure}
\includegraphics[width=0.45\textwidth]{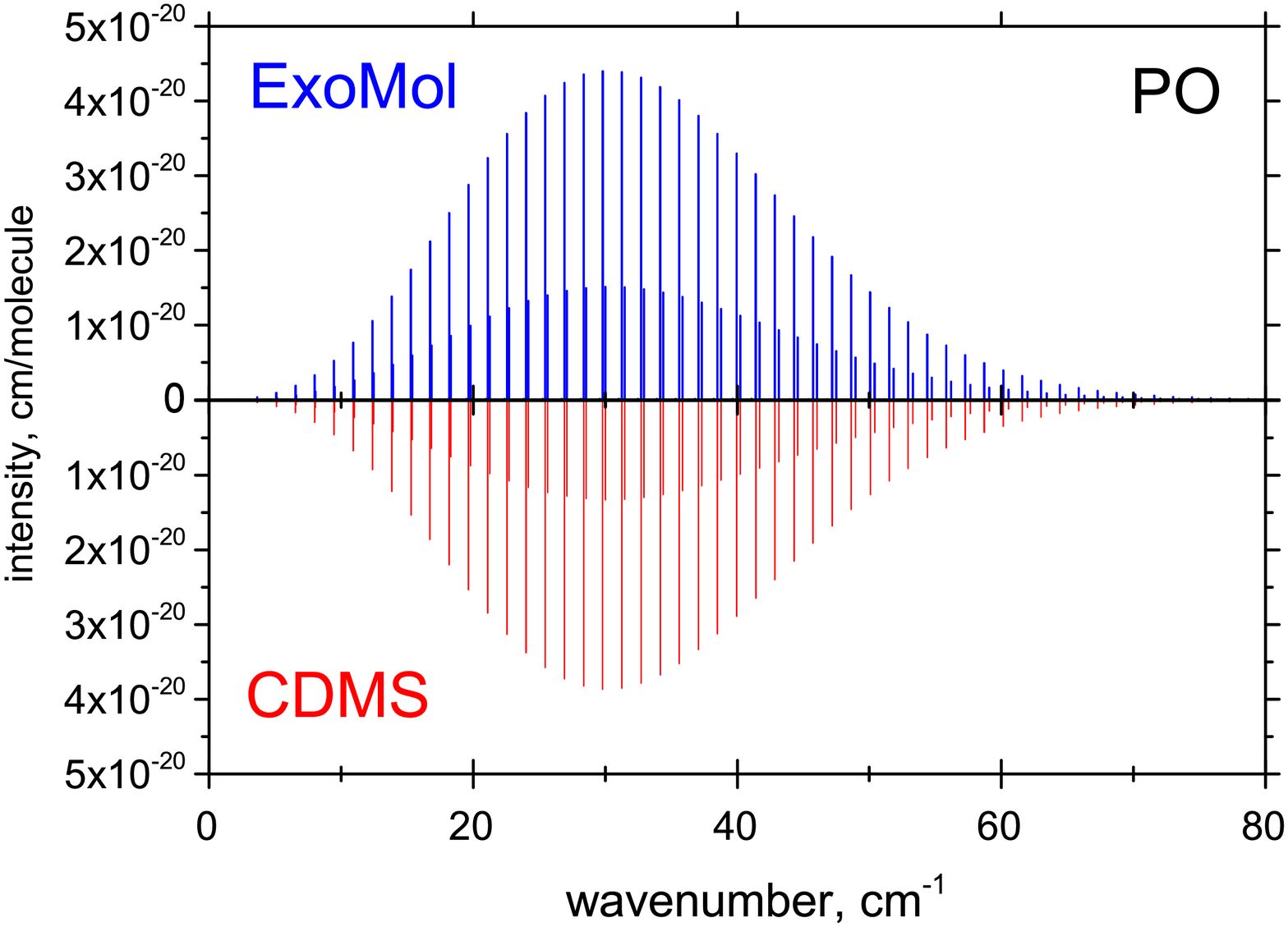}
\includegraphics[width=0.45\textwidth]{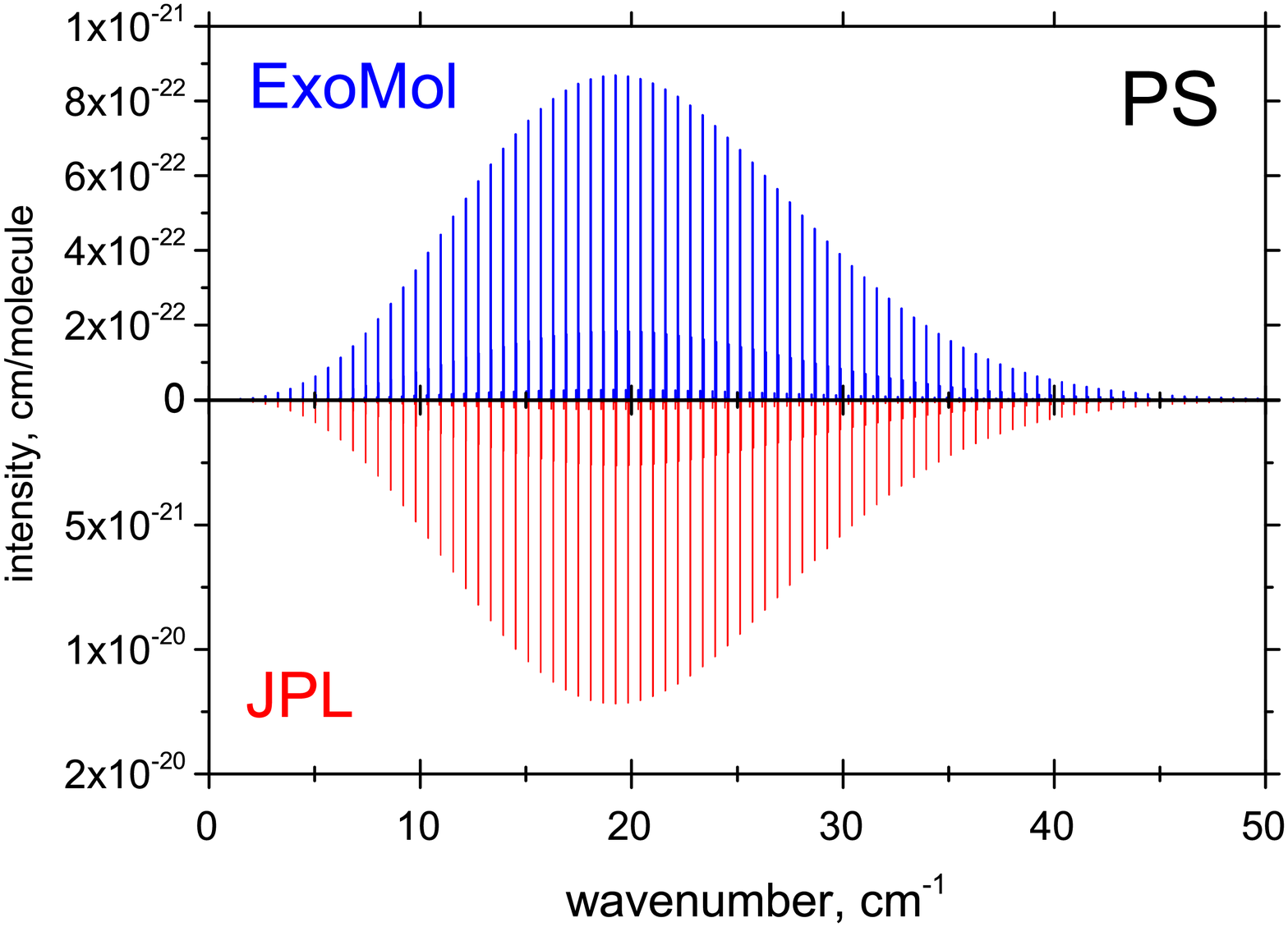}
\caption{Comparison of pure PO (left) and PS (right) rotational lines ($T$ = 298 K) with those given in CDMS  \citep{CDMS} and JPL
\protect\citep{jpl} databases, respectively. }
\label{f:PO_rotspec}

\end{figure}

Figure~\ref{f:PO_rotspec} gives a comparison of our pure rotational
spectrum  with that given by CDMS \citep{CDMS} at $T=$ 298~K.
Hyperfine lines given in CDMS have been convolved for the comparison.
The two spectra are in excellent
agreement apart from a small difference in line intensities.
This slight discrepancy arises from different values for the dipole moment:
CMDS used the experimental equilibrium value 1.88~D by \citet{88KaYaSa.PO}, while our
\ai\ dipole is 1.998~D at $r_{\rm e}$.

\begin{figure}
\includegraphics[width=0.45\textwidth]{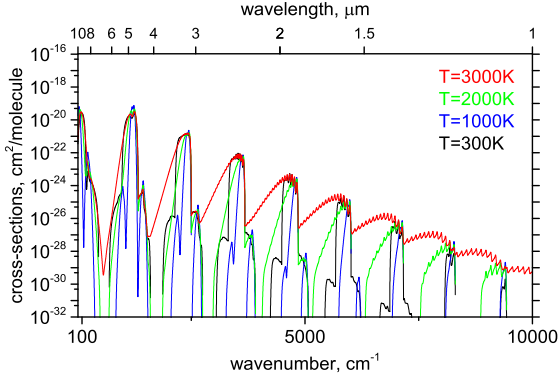}
\caption{PO absorption spectrum at $T=$ 300 (bottom), 1000, 2000  and 3000 (top)~K,
presented with cross-sections on a logarithmic scale. A Gaussian profile with HWHM=10~\cm\ was used. }
\label{f:PO_abs}
\end{figure}

Figure~\ref{f:PO_abs} shows our predicted, temperature-dependent absorption spectrum for PO.

\subsection{PS $X$~$^{2}\Pi$ state}

Figure~\ref{f:PO_rotspec} shows a comparison of the rotational spectrum ($T =$ 298~K) of PS simulated using our line list with that from the JPL database \citep{jpl}. The latter is also a simulated spectrum where a generic dipole value of 2~D was used; therefore, the JPL's intensity scale is arbitrary. We
recommmend that JPL's intensities are rescaled using value for the
dipole.

\begin{figure}
\includegraphics[width=9cm]{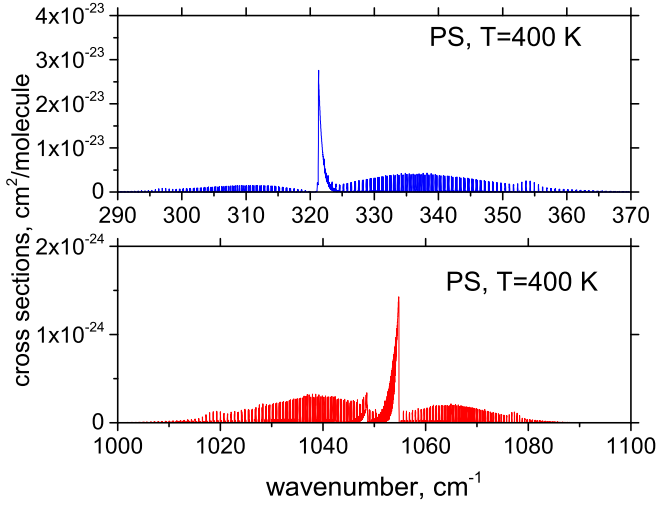}
\caption{The $Q$-branches in the pure rotational and  fundamental absorption
bands of PS at $T=$ 400~K. A Gaussian profile with HWHM of 0.02~\cm\ was used.   }
\label{f:PS_spec}
\end{figure}

The $Q$-branches from pure rotational and fundamental bands are illustrated in Fig.~\ref{f:PS_spec}, where the absorption spectra of PS
at $T=$ 400~K are shown. These sharp features are important to astronomers because they are easily identifiable; however,
the transitions themselves are relatively weak.

\begin{figure}
\centering
\includegraphics[width=8cm]{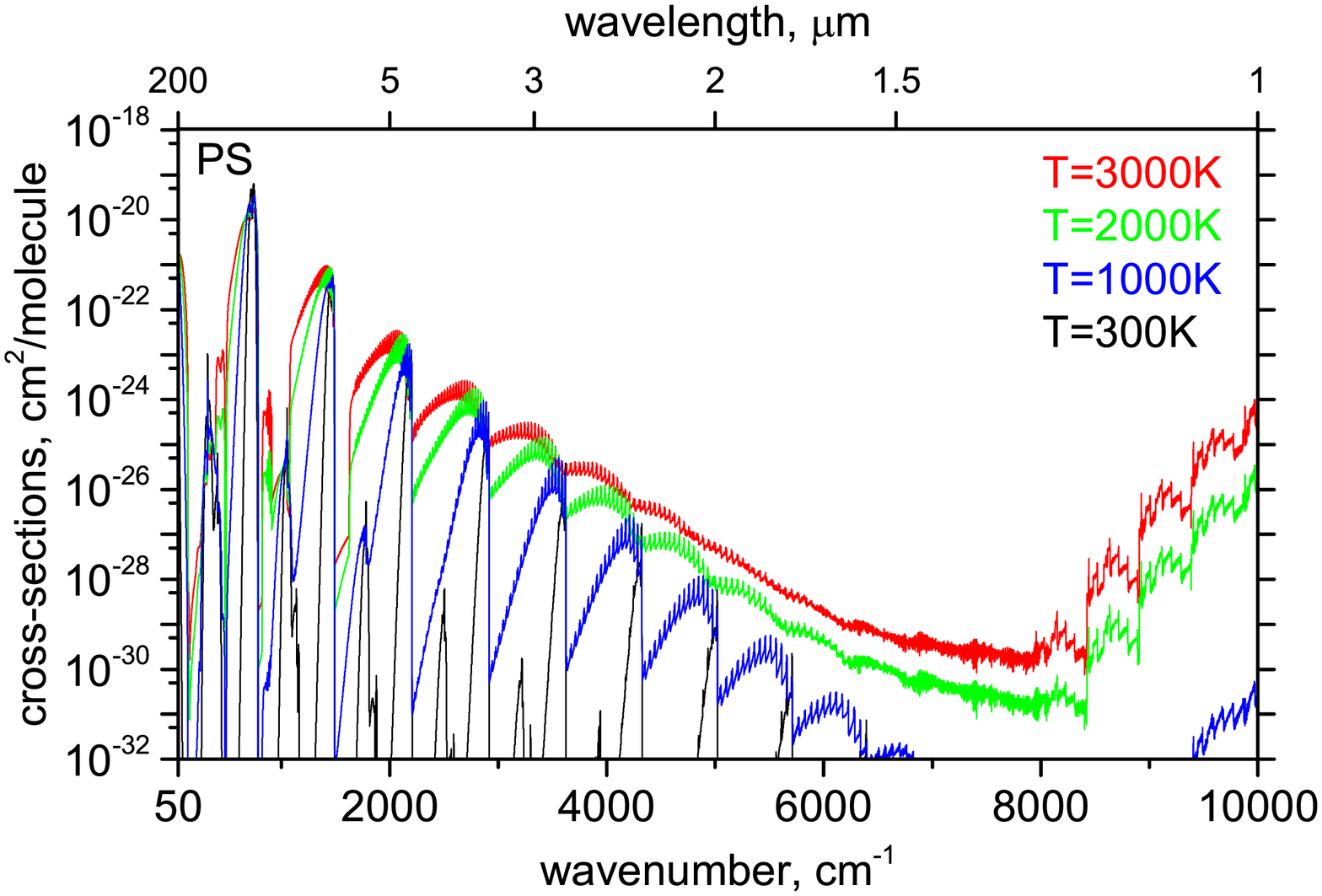}
\includegraphics[width=8cm]{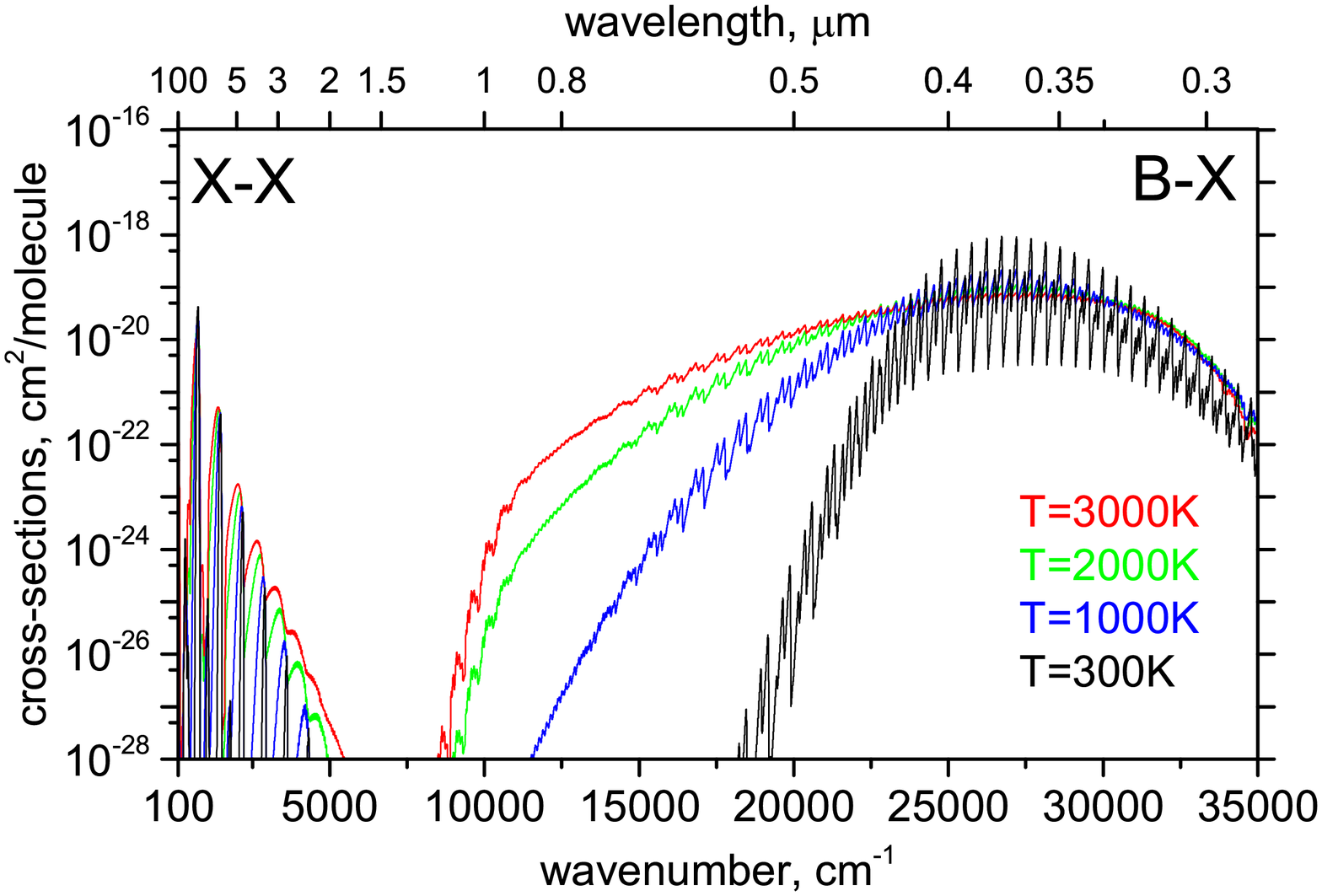}

\caption{Absorption spectrum for the ground state of PS as a function of temperature
($T=$ 300~K, 1000~K, 2000~K and 3000~K).   Gaussian profiles with HWHM=1~\cm\ (left panel) and 10~\cm (right panel) were used.  }
\label{f:PS_abs:T}
\end{figure}

Figure~\ref{f:PS_abs:T} displays the temperature dependent absorption spectrum of PS.
The spectra are prominent in the infrared as expected for rovibrational
transitions within the ground electronic state.
The fundamental band (${v}=1$$\leftarrow$0) at about 13.6 $\mu$m
corresponds to the highest peak in the spectra ($\sim1\times10^{-19}\:\mathrm{cm^{2}/molecule}$).
The vibrational overtones at shorter wavelengths than the fundamental
band are progressively weaker.

Cross-sections were calculated for transitions corresponding
to the \textit{allowed} rovibrational transitions
within the $X$~$^{2}\Pi$, $a$~$^{4}\Pi$ and $B$~$^{2}\Pi$ terms
and for $B$~$^{2}\Pi$--$X$~$^{2}\Pi$ transitions. The $B$~$^{2}\Pi$--$X$~$^{2}\Pi$ band
peaks in the UV region of the electromagnetic spectrum as expected.

Our predicted temperature-dependent  $B$~$^{2}\Pi$--$X$~$^{2}\Pi$ absorption
spectrum is shown in Fig.~\ref{f:PS_abs:T}. At low temperatures it shows
sharp features in the 3000 -- 4000 \AA\ region. These features gradually diminish at
higher temperatures. Although the $B$~$^{2}\Pi$--$X$~$^{2}\Pi$
band remains featureless at high temperatures, it could possibly be
used as a tracer for PS in low temperature astronomical environments.

\begin{figure}
\centering
\includegraphics[width=8cm]{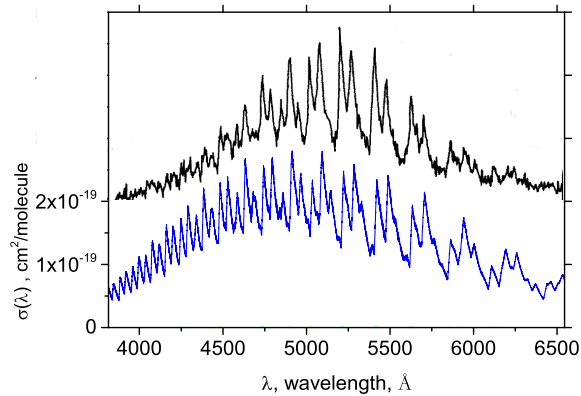}
\caption{Predicted $B$--$X$ emission spectrum of PS at $T=2000$~K (lower curve) with a Gaussian profile with HWHM=50~\cm,
compared to PS chemiluminescence  from the reaction Cs$+$PSCl$_3$ by \protect\citet{87LiBaWr.PS} (upper curve). The cross section scale refers to our
computed values only. }
\label{f:2000K:emiss}
\end{figure}

Figure \ref{f:2000K:emiss} shows a comparison of a synthetic ($T=2000$~K) $B$--$X$ emission spectrum with a chemiluminescence spectrum from the reaction Cs$+$PSCl$_3$ observed by \citet{87LiBaWr.PS}.

\section{Discussion and Conclusion}

The line lists for PO ($X$~$^{2}\Pi$) and PS ($X$~$^{2}\Pi$, $B$~$^{2}\Pi$,
and $a$~$^{4}\Pi$) are the most comprehensive to date. The PO line list, covering the $X$~$^{2}\Pi$ state,
should be sufficient for modelling infrared spectra for this species
at long wavelengths. The PS line list has a larger coverage, up to
37,000~\cm\ and considers two low-lying excited electronic states,
$B$~$^{2}\Pi$ and $a$~$^{4}\Pi$ in addition to the ground $X$~$^{2}\Pi$ state. This
uses the refined curves for the $X$ and $B$ states, but with the addition
of the unrefined \ai\ curves for the $a$~$^{4}\Pi$ electronic
state as relevant experimental data was unavailable.  This line list is
designed to aid the identification of spectral features of PS at short
wavelengths, although a more accurate line list will be necessary for
a full spectral analysis. Our line list shows a strong banded
structure for absorption at 300~K in the 3500 \AA\ region. However, at
higher temperatures the absorption becomes a very broad feature with
little to no structure suggesting that observation of the
$B$~$^{2}\Pi$--$X$~$^{2}\Pi$ is unlikely to be useful for detecting PS in
hot environments. The $Q$-branch transitions from the spin-orbit split
components can be used as a diagnostic for PS in observational
spectra due to the characteristically sharp peak in both
absorption and emission.  However, the transitions themselves are
relatively weak, and so sensitive detectors would be required.

It is difficult to provide specific uncertainties for theoretical calculations.
However, our fits to the experimental transition frequencies give some
insight into the accuracy of the line positions. For PO this is 0.001 \cm\ for
pure rotational transitions and 0.05 \cm\ for transitions involving
vibrational excitation. These estimates cover the range were
experimental data is available ($v \leq 11$ and $J \leq 22.5$); for
values outside these ranges the uncertainty is expected to grow
approximately linearly with $v$ and quadratically with $J$.
Comparison with the rather uncertain experimentally determined dipole suggest
that our computed intensities for PO are unlikely to be accurate to better
than 10\%, and are possibly worse than this. For PS there is less high-quality
experimental data for line positions available. For transitions within the $X$ state, pure
rotational transitions are accurate to about 0.006 \cm\ for $J \leq 39.5$;
the accuracy of the vibrational transition frequencies are limited by
a lack of high accuracy measurements and are unlikely to good to better
than 0.2 \cm. In the absence of any experimental determinations
we can only estimate the uncertainty of vibration-rotation transition
intensities as being between 10 and 20~\%. The vibronic line list for PS
is considerably less accurate with transition frequencies unlikely to be
better than 0.5 \cm\ and intensities only reliable to about $\pm 50$\%.

These line lists, which we call the POPS line lists, can be downloaded from the \url{http://cdsarc.u-strasbg.fr}  or from \url{www.exomol.com}.

So far phosphorus mononitride (PN) \citep{jt590} and
phosphine ($\mathrm{PH_{3}}$) \citep{jt592} are the only phosphorus-bearing molecules whose line lists have been computed as part of the
\textsc{ExoMol} project.  PN was the first P-bearing molecule to be astronomically
observed by \citet{87TuBaxx.PN} and
\citet{87Zixxxx.PN} in Orion KL, Sagittarius B2 and W51; PH$_3$
was discovered in Jupiter and Saturn from Voyager data in 1975 \citep{75BrLeRa,76RiWaSm},
and recently detected around the C-rich AGB star, IRC$+$10216 \citep{08AgCePa}. This paper
presents line lists for PO and PS. A line list for PH is currently under construction
while an empirical line list for CP, which can also be downloaded from the ExoMol website, was provided by \citet{14RaBrWe.CP}.

\section{Acknowledgements}

This work was supported by the ERC under the Advanced Investigator Project
267219. We also acknowledge the networking support by the COST Action CM1405 MOLIM.
This work made extensive use of the Legion HPC at UCL.

\bibliographystyle{mnras}

\label{lastpage}

\end{document}